\documentclass[10pt]{book}
\usepackage{array}
\usepackage{setspace}
\usepackage{color}
\usepackage{amssymb}
\usepackage{amsmath}
\usepackage{epsfig}
\usepackage[authoryear]{natbib}
\newtheorem{corollary}{Corollary}[section]
\newtheorem{theorem}{Theorem}[section]
\newtheorem{lemma}{Lemma}[section]

\begin{document}

\setcounter{page}{1}
%\begin{document}
%\pagestyle{fancy}
\renewcommand{\baselinestretch}{1.2}
%\lhead[\fancyplain{} \leftmark]{}
%\chead[]{}
%\rhead[]{\fancyplain{}\rightmark}
%\cfoot{}
%\headrulewidth=0pt
\markright{
%\hbox{\footnotesize\rm Statistica Sinica
%{\footnotesize\bf ??}(200?), 000-000}\hfill
}
\markboth{\hfill{\footnotesize\rm ABHYUDAY MANDAL, JIE YANG AND DIBYEN MAJUMDAR
}\hfill}
{\hfill {\footnotesize\rm OPTIMAL DESIGN FOR BINARY RESPONSE} \hfill}
\renewcommand{\thefootnote}{}
$\ $\par
\fontsize{10.95}{14pt plus.8pt minus .6pt}\selectfont
\vspace{0.8pc}
\centerline{\large\bf OPTIMAL DESIGNS FOR TWO-LEVEL FACTORIAL EXPERIMENTS}
\vspace{2pt}
\centerline{\large\bf WITH BINARY RESPONSE}
\vspace{.4cm}
\centerline{Abhyuday Mandal$^{1}$, Jie Yang$^{2}$ and Dibyen Majumdar$^{2}$}
\vspace{.4cm}
\centerline{\it  $^1$University of Georgia and $^2$University of Illinois}
\vspace{.55cm}
\fontsize{9}{11.5pt plus.8pt minus .6pt}\selectfont

\begin{quotation}
\begin{center} {\bf{ \it Abstract:}}\end{center}
\bigskip \noindent  We consider the problem of obtaining locally D-optimal designs for factorial experiments with qualitative factors at two levels each with binary response. Our focus is primarily on the $2^2$ experiment. In this paper, we derive analytic results for some special cases and indicate how to handle the general case. The performance of the uniform design in examined and we show that this design is highly efficient in general. For the general $2^{k}$ case we show that the uniform design has a \textit{maximin} property.
\par

\vspace{9pt} \noindent {\it Key words and phrases:} Generalized linear model, full factorial design, D-optimality, information matrix, uniform design, cylindrical algebraic decomposition.
\par
\end{quotation}\par

\fontsize{10.95}{14pt plus.8pt minus .6pt}\selectfont
\setcounter{chapter}{1}
\setcounter{equation}{0} %-1
\noindent {\bf 1. Introduction}

\bigskip\noindent The goal of many scientific and industrial experiments is to study a process that depends on several qualitative factors. We will focus on the design of those experiments where the response is binary. If the response was quantitative and a linear model was appropriate, then the design of the experiment is informed by the extensive literature on factorial experiments. On the other hand if the factors were quantitative and the response was binary, the literature on optimal design of generalized linear models in the approximate theory setup could be used. The goal of our work is to initiate the optimal design theory for factorial experiments with binary response.

\bigskip\noindent Specific examples of experiments of the type we are interested in are available in different areas of application. Smith (1932) describes a bioassay for an anti-pneumococcus serum where the explanatory variable is doses of the serum. Mice infected with pneumococcus are injected with different doses of the serum and the response is survival (or not) beyond seven days. Hamada and Nelder (1997) discussed the advantages of using a generalized linear model for discrete responses instead of linearizing the response to obtain an approximate linear model. They examined an industrial experiment on windshield molding performed at an IIT Thompson plant that was originally reported by Martin, Parker and Zenick (1987). There were four factors each at two levels and the response was whether the part was \textit{good} or not.  A $2^{4-1}$ fractional factorial design was used with $1000$ runs at each experimental condition. Other examples include a seed gemination experiment described in Crowder (1978), a sperm survival experiment in Myers, Montgomery and Vining (2002) and a designed experiment on the reproduction of plum trees reported by Hoblyn and Palmer (1934).

\bigskip\noindent We assume that the process under study may be adequately described by a generalized linear model. While the theory we develop will work for any link function, in examples and simulations we focus on the logit, probit, log-log, and complementary log-log links. The optimal designs will be obtained using the D-criterion that maximizes the determinant of the inverse of the asymptotic covariance matrix of the estimators (the information matrix). In order to overcome the difficulty posed by the dependence of the design optimality criterion on the unknown parameters, we use the local optimality approach of Chernoff (1953) where the parameters are replaced by assumed values. We refer the reader to the paper by Khuri, Mukherjee, Sinha and Ghosh (2006) for details of theory of designs for generalized linear models.

\bigskip\noindent We will assume that every factor is at two levels, a setup of particular interest in screening experiments, and that (for an experiment with $k$ factors) we are interested in a \textit{complete} $2^{k}$ experiment, i.e., the design may be supported on all $2^{k}$ points. The model we choose may include a subset of all main effects and interactions. If we assume that the total number of observations is held fixed, then the design problem is to determine the proportion of observations allocated to each of the $2^{k}$ design points. It may be noted that if the response follows a standard linear model, then it follows from the results of Kiefer (1975) that the design which is \textit{uniform} on the $2^{k}$ design points is universally optimal. For the problem restated in terms of weighing design, Rao (1971) gave the optimality of the uniform design in terms of minimizing variances of each of the parameter estimators. It may be noted that the uniform design is an \textit{orthogonal array} (Rao (1947)).

\bigskip\noindent In this initial study, we focus primarily on the complete $2^{2}$ experiment where the response is binary. While we do not find analytic solutions for D-optimal design for the general $2^{2}$ experiment, we obtain characterizations for several special cases. For the general $2^{2}$ experiment we indicate how a solution may be obtained by \textit{Cylindrical Algebraic Decomposition} (CAD). We also examine uniform designs and show that these are highly efficient in general.

\bigskip\noindent For the general $2^{k}$ experiment we show that the uniform design is maximin D-optimal design, i.e., a design that maximizes a lower bound of the D-criterion.

\bigskip\noindent In section 2 we give the preliminary setup. Results for the $2^{2}$ experiment are proved in section 3. In section 4 we study robustness of the uniform design, and in section 5 we consider the general $2^{k}$ experiment. Some concluding remarks are given in section~6. Proofs are relegated to the appendix and details are also available under the supplementary materials.

\setcounter{chapter}{2}
\setcounter{section}{1}
\setcounter{equation}{0} %-1
\bigskip \noindent {\bf 2. Preliminary Setup}

\bigskip\noindent Consider a $2^{k}$ experiment, i.e., an experiment with $k$ explanatory variables at $2$ levels each. Suppose $n_{i}$ units are allocated to the $ i$th experimental condition such that $n_{i}\geqslant 0,$ $i=1,\ldots,2^{k}$, and $n_{1}+\cdots+n_{2^{k}}=n$. We suppose that $n$ is fixed and consider the problem of determining the ``optimal" $n_{i}$'s. \ In fact, we write our optimality criterion in terms of the proportions:\vspace{-.1in}%
\begin{equation*}
p_{i}=n_{i}/n,\text{ }i=1,\ldots,2^{k}
\end{equation*}
and determine the ``optimal" $p_{i}$'s. (Since $n_{i}$'s are integers, an optimal design obtained in this fashion may not be ``feasible" - an issue we will not deal with, except to say that a feasible solution ``near" an optimal solution is expected to be ``nearly optimal").

\bigskip\noindent Suppose $\eta $ is the linear predictor that involves main effects and interactions which are assumed to be in the model. Our main focus in this initial research on the topic will be the $2^{2}$ experiment with main-effect model, in which case $ \eta =\beta _{0}+\beta _{1}x_{1}+\beta _{2}x_{2}$ and $\beta =\left( \beta _{0},\beta _{1},\beta _{2}\right) ^{\prime }$. In the framework of generalized linear models, the response $Y$ is linked to the linear predictor by the link function $g$: $E\left( y\right) =\mu ,$ $\eta =g\left( \mu \right) $ (McCullagh and Nelder (1989)). For a binary response, the commonly used link functions are logit, probit, log-log, and complimentary log-log links.

\bigskip\noindent The maximum likelihood estimator of $\beta $ has an asymptotic covariance matrix (Khuri, Mukherjee, Sinha and Ghosh (2006)) that is the inverse of $ nX^{\prime }WX$, where $W=diag\left( w_{1}p_{1},...,w_{2^{k}}p_{2^{k}}\right) ,$ $w_{i}=\left( \frac{d\mu _{i}}{ d\eta _{i}}\right) ^{2}/\left( \mu _{i}(1-\mu _{i})\right) \geq 0$ and $X$ is the ``design matrix''. For a main-effect $2^{2}$ experiment, for instance, $X = ((1, 1, 1, 1)'$, $(1, 1, -1, -1)'$, $(1, -1, 1, -1)')$. The matrix $X^{\prime }WX$ may be viewed as the \textit{per-observation} information matrix. The {\it D-optimality} criterion maximizes the determinant $\left\vert X^{\prime }WX\right\vert .$

\setcounter{chapter}{3}
\setcounter{section}{1}
\setcounter{equation}{0} %-1
\bigskip \noindent {\bf 3. D-Optimal $2^2$ Designs}

\noindent If $k=2$, for a main-effect plan, the asymptotic information matrix is proportional to $X'WX$.
It can be shown that $|X^\prime WX|$ can be written as (except for the constant $16$):
\begin{eqnarray}\label{maineq}
det(\mathbf{w},\mathbf{p}) = G(\mathbf{p})&=&p_2w_2\cdot p_3w_3\cdot p_4w_4 + p_1w_1\cdot p_3w_3\cdot p_4w_4\nonumber \\
                  &+&p_1w_1\cdot p_2w_2\cdot p_4w_4 + p_1w_1\cdot p_2w_2\cdot p_3w_3
\end{eqnarray}
where ${\mathbf w}=(w_1,w_2,w_3,w_4)'$ and ${\mathbf p}=(p_1,p_2,p_3,p_4)'$. In this section, we will consider the problem of maximizing $G(\mathbf{p})$ over all vectors $\mathbf{p}$ with $p_i \geq 0$ and $\sum_ip_i=1$.

\vspace{0.5cm}
\setcounter{section}{1}
\noindent
{\bf 3.1 Analytic solutions to special cases}

\bigskip\noindent It follows from Kiefer (1975) that if all the $w_i$'s are equal then the uniform design ($p_1 = p_2 = p_3 = p_4 = 1/4$)
is D-optimal. If one and only one of the $w_i$'s is zero, then the optimal design is uniform over the design points that correspond to the nonzero $w_i$'s, and if two or more $w_i$'s are zero, then $G(\mathbf{p})\equiv 0$. From now on, we assume $w_i>0$, $i=1, 2, 3, 4$. Define $L=G/(w_1w_2w_3w_4)$ and
$v_i = 1/w_i,\ i=1,2,3,4.$
The maximization problem (\ref{maineq}) can be rewritten as maximizing
\begin{equation}\label{simplifiedprob}
L({\mathbf p})=v_4p_1p_2p_3+v_3p_1p_2p_4+v_2p_1p_3p_4+v_1p_2p_3p_4.
\end{equation}
Although the objective function (\ref{simplifiedprob}) is elegant, an analytic solution with general $v_i>0$ is not available.
In this subsection, analytic solutions are obtained for some special cases.
%The proofs are relegated to the appendix.

\begin{theorem}\label{thm:1} $L(\mathbf{p})$ has a unique maximum at ${\mathbf p} = (0, 1/3, 1/3, 1/3)$ if and only if $v_1\geq v_2+v_3+v_4$.
\end{theorem}

\noindent Note that this does not correspond to a complete $2^2$ experiment, rather it corresponds to a design supported only on three points, which is saturated for the main effects plan $\eta =\beta_0 + \beta_1x_{1} + \beta_2 x_{2}$. For the logit link function, $0\leq w_i\leq 0.25$.
Computations under the simulation condition that $w_i$ iid $\sim$ uniform(0,0.25) show that the chance of obtaining
%``$2\max_i v_i \geq v_1 + v_2 + v_3 + v_4$"
a saturated solution is 48\%. For other link functions, the chances are similar.

\begin{lemma}\label{lem:1}
If $v_1>v_2$, then any solution to the maximization problem of (\ref{simplifiedprob}) must satisfy $p_1\leq p_2$; if $v_1=v_2$, then any solution must satisfy $p_1=p_2$.
\end{lemma}

\begin{theorem}\label{thm:2}
Suppose $v_1 \ge v_2$, $v_3=v_4=v$, and $v_1<v_2+2v$. Then the solution maximizing (\ref{simplifiedprob}) is
\begin{equation}p_1=\frac{1}{2}-\frac{v_1-v_2+4v}{2(-2\delta+D)},\> p_2=\frac{1}{2}+\frac{v_1-v_2-4v}{2(-2\delta+D)},\>
p_3=p_4\> =\> \frac{2v}{-2\delta+D}\label{innersolution}
\end{equation}
with $L=2v^2\left(\delta^2+4v_1v_2-\delta D\right)/(-2\delta+D)^3$, where $\delta=v_1+v_2-4v$ and $D=\sqrt{\delta^2+12v_1v_2}$.
\end{theorem}

\begin{corollary}Suppose $v_2=v_3=v_4=v$ and $v_1 < 3v$.  Then the solution maximizing (\ref{simplifiedprob}) is
\[
    p_1=\frac{3v - v_1}{9v-v_1},\> p_2=p_3=p_4=\frac{2v}{9v-v_1}
\]
with the maximum $L=4v^3/(9v-v_1)^2$.
\end{corollary}

\begin{corollary}Suppose $v_1=v_2=u$, $v_3=v_4=v$, and $u>v$. Then the solution maximizing (\ref{simplifiedprob}) is
\[
    p_1=p_2=\frac{2u-v-d}{6(u-v)},\> p_3=p_4=\frac{u-2v+d}{6(u-v)}%\label{caseiii}
\]
with the maximum $L=\left.(2u-v-d)(u-2v+d)(u+v+d)\right/\left[108(u-v)^2\right]$, where $d = \sqrt{u^2 - uv + v^2}$.
\end{corollary}

\bigskip \noindent Theorem~\ref{thm:1} reveals that the D-optimal design is saturated if and only if $2\max_i v_i \geq v_1 + v_2 + v_3 + v_4$. We call it the {\it saturation condition}. In terms of $w$'s, it is
\begin{equation}\label{boundarycondition}
2/\min\{w_1,w_2,w_3,w_4\} \geq 1/w_1 + 1/w_2 + 1/w_3 + 1/w_4.
\end{equation}

\bigskip\noindent Theorem~\ref{thm:3} examines this condition in terms of the $\beta$'s.

\bigskip \noindent
\begin{theorem}\label{thm:3}
For the logit link, the saturation condition is true if and only if
\[\beta_0 \neq 0,\>\>\>\>\>
 |\beta_1| > \frac{1}{2}\log\left(\frac{e^{2|\beta_0|}+1}{e^{2|\beta_0|}-1}\right),
 \>\>\mbox{ and }
\]
\[
 |\beta_2| \geq \log\left( \frac{2e^{|\beta_0|+|\beta_1|}+\sqrt{\left(e^{4|\beta_0|}-1\right)\left(e^{4|\beta_1|}-1\right)}} {\left(e^{2|\beta_0|}-1\right)\left(e^{2|\beta_1|}-1\right)-2}\right).
\]
\end{theorem}

\begin{tabular}[h]{c}
\includegraphics[height=2.25in,width=2.25in,angle=0]{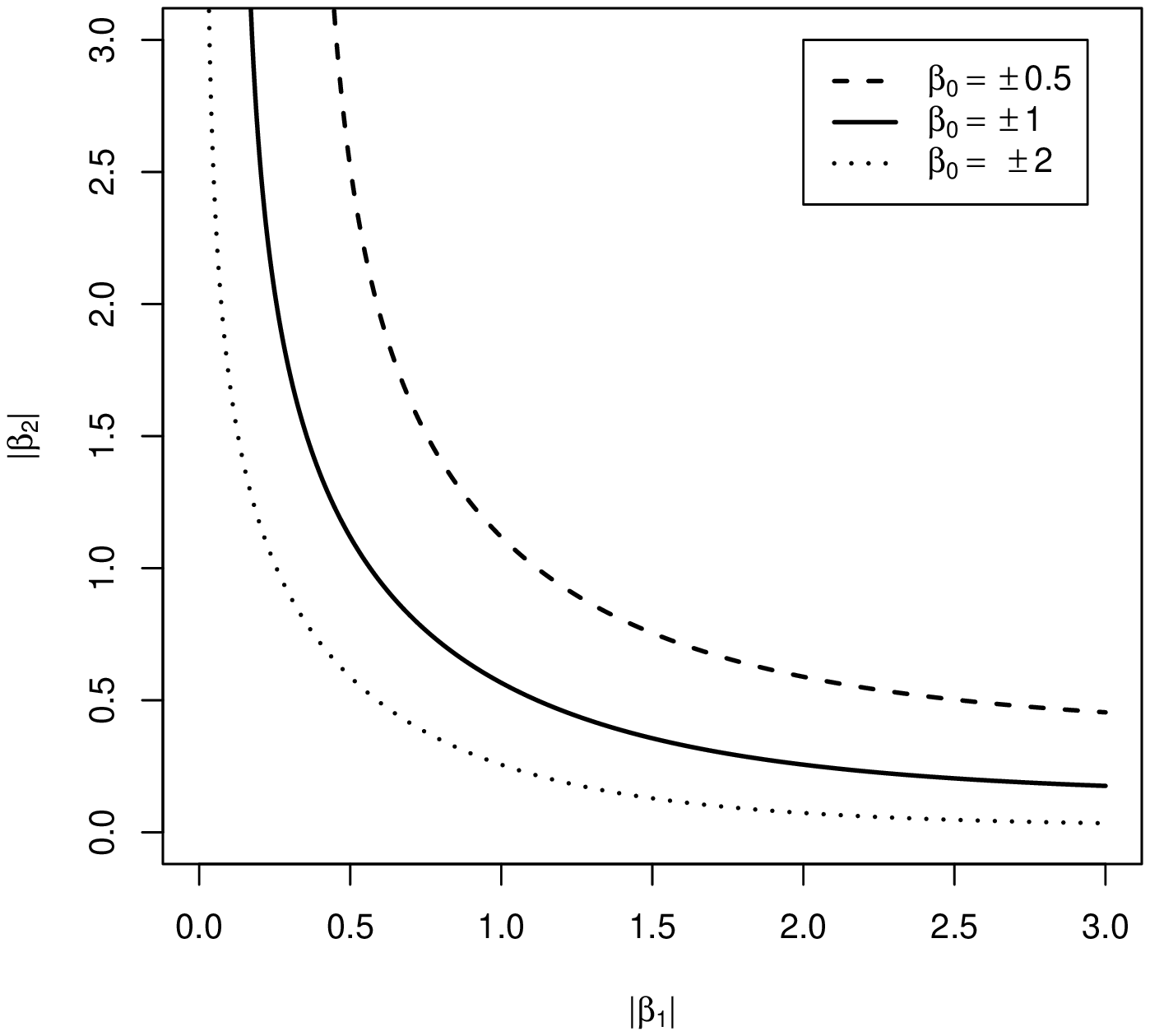} \\
Figure~1: Lower boundary of the region satisfying the saturation condition \\
\end{tabular}

\bigskip \noindent Figure~1 shows how the region satisfying the saturation condition changes with $\beta_0$. For fixed $\beta_0$, a pair ($\beta_1$, $\beta_2$) satisfies the saturation condition if and only if the corresponding point in Figure~1 is above the curve labelled by $\beta_0$.

\bigskip \noindent For logit link, $\beta_0,\beta_1,\beta_2$ are symmetric with respect to $w_1,w_2,w_3,w_4$.
If we permute $\beta_0,\beta_1,\beta_2$, we only need to change the order of $w_1,w_2,w_3$ accordingly. It won't affect whether or not the saturation condition is satisfied.  Therefore,
we can get parallel results to Theorem~\ref{thm:3} in any order of $\beta_0,\beta_1,\beta_2$.

\setcounter{section}{2}
\vspace{0.5cm}

\bigskip \noindent {\bf 3.2 Exact solution using computer-aid optimization}

\bigskip \noindent For given values of $v_1,v_2,v_3,v_4$, depending on the computational resources, one can always try to apply some numerical searching algorithm such as Nelder-Mead, quasi-Newton, conjugate-gradient, or simply a grid search. However, such solutions will not be very accurate in general.

\bigskip \noindent For general cases other than the special ones discussed above, one may use {\it cylindrical algebraic decomposition} (CAD) (Fotiou et al. (2005)) to find an exact solution maximizing (\ref{simplifiedprob}). For given $v_1, v_2, v_3, v_4$, the CAD algorithm partitions the feasible domain of $(L, p_1, p_2, p_3)$ into a finite union of disjoint homogeneous cells in terms of polynomial constrains. The cell with greatest $L$ provides us a solution. However, it is difficult, if not computationally infeasible, to get explicit formula involving general $v_1,v_2,v_3,v_4$.

\bigskip \noindent Besides the CAD approach, one may also use the Lagrange multipliers or the Karush-Kuhn-Tucker (KKT) conditions (Karush (1939), Kuhn and Tucker (1951)) to find all the local extrema of $L$, and then pick the largest one. For general $v_1,v_2,v_3,v_4$, this approach leads to two polynomial equations of order $4$ each. No explicit formula involving general $v_1,v_2,v_3,v_4$ is available.

\setcounter{section}{3}
%\vspace{1.5cm}

\bigskip \noindent {\bf 3.3 Analytic approximate solution}

\bigskip \noindent For the optimization problem (\ref{simplifiedprob}), we do not find analytic solutions for the general case. In this section, we will propose an analytic approximate solution. To simplify notations, write $L[v_1,v_2,v_3,v_4]$ for $\max_{\mathbf p}L$, given $v_1, v_2, v_3, v_4$. For example, Theorem~\ref{thm:2} corresponds  to $L[v_1,v_2,v,v]$. Without any loss of generality, we assume $v_1<v_2<v_3<v_4$ and $v_4 < v_1+v_2+v_3$.
Define  \vspace{-.1in}
\[
L_{34}=L[v_1,v_2,(v_3+v_4)/2,(v_3+v_4)/2]
\]
and $L_{12}$, $L_{13}$, $L_{14}$, $L_{23}$, $L_{24}$ accordingly.
The strategy is to
use
$\max\{L_{12}, L_{23}, L_{34}\}$
to approximate $\max_{\mathbf p} L$ based on the theorem as follows.

\begin{theorem}\label{thm:4} Assume $v_1<v_2<v_3<v_4$ and $v_4 < v_1+v_2+v_3$.  Then
\begin{eqnarray*}
 {\max}\{L_{13},L_{14},L_{24}\} &\leq&  \max\{L_{12},L_{23},L_{34}\},\\
 {\max}_{\mathbf p} L  - \max\{L_{12},L_{23},L_{34}\} &\leq& \min\left\{\frac{v_2-v_1}{216}, \> \frac{v_3-v_2}{96\sqrt{3}},\> \frac{v_4-v_3}{54} \right\}.
\end{eqnarray*}
\end{theorem}

\bigskip \noindent We call the best $\mathbf{p}$ among the solutions to $L_{12}$, $L_{23}$, or $L_{34}$ the {\it analytic approximate solution}, and denote it by $\mathbf{p_a}$. Then $L(\mathbf{p_a}) = \max\{L_{12}, L_{23}, L_{34}\}$. Theorem~\ref{thm:4} provides a theoretical upper bound for the difference ${\max}_{\mathbf p} L  - L(\mathbf{p_a})$.

\bigskip \noindent To see how our approximation works numerically, we randomly selected 1000 $\mathbf{w} = (w_1,$ $w_2,$ $w_3,$ $w_4)$ with $0.05 \leq w_i \leq 0.25$. For each randomly chosen $\mathbf{w}$, we calculate the optimal $\mathbf{p}$ using CAD and denote it by  $\mathbf{p_o}$. We also calculate the analytic solution based on Theorems \ref{thm:1}, \ref{thm:2}, or \ref{thm:4} and denote it by $\mathbf{p_*}$. If the saturation condition (\ref{boundarycondition}) is satisfied or $w_i = w_j$ for some $i\neq j$, then $\mathbf{p_*} = \mathbf{p_o}$. Otherwise, the analytic approximate solution based on Theorem~\ref{thm:4} is applied, and $\mathbf{p_*}=\mathbf{p_a}$. Then we calculate the determinant of the information matrix and denote it by $D_o$ and $D_*$ for $\mathbf{p_o}$ and $\mathbf{p_*}$, respectively. In Figure~2, we plot the relative loss $\left.\left(D_o^{1/3} - D_*^{1/3}\right)\right/D_o^{1/3}$ versus $D_o^{1/3}$.  Numerical results show that 96\% of the relative losses are less than 0.03\% and the maximum relative loss is about 0.085\%. So the analytic approximation solution works very well.

\bigskip \noindent Since $v_i$'s depend on the assumed values, they are in general not quantified accurately at the planning stage. The results in this section show that setting some $v_i$'s to be equal is not a bad strategy. The analytic approximation is also potentially useful for future theoretical research in this area.

\begin{center}
\begin{tabular}{c}
\includegraphics[height=2.5in,width=2.5in,angle=0]{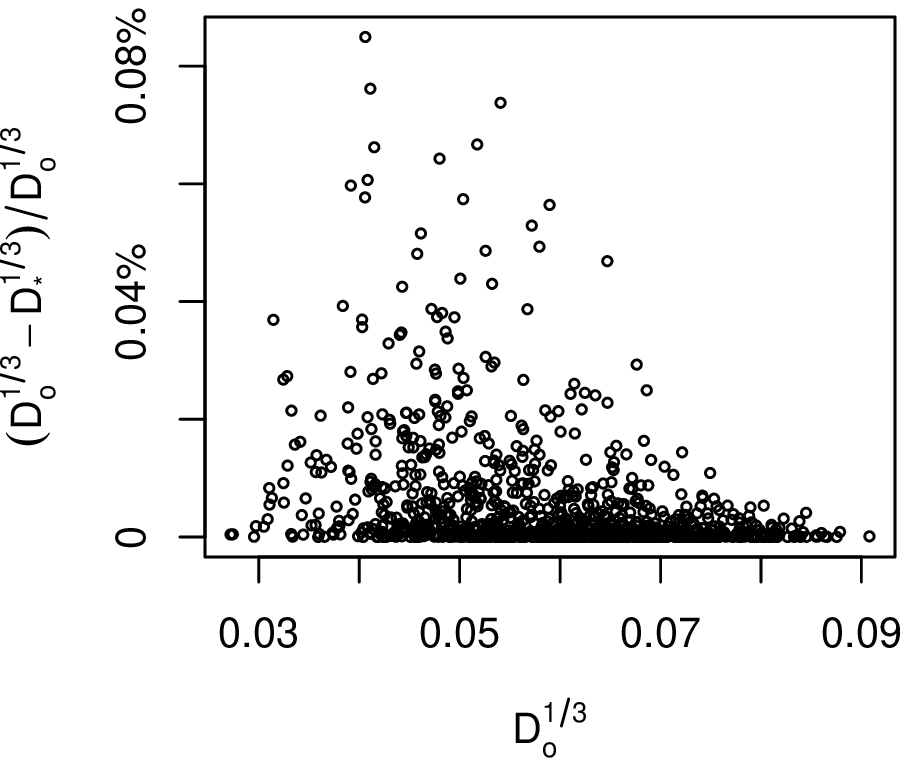}\\
Figure~2: Efficiency of the approximate solutions \\
\end{tabular}
\end{center}

\setcounter{chapter}{4}
\setcounter{section}{1}
\setcounter{table}{0}
\setcounter{equation}{0} %-1
\bigskip \noindent {\bf 4. Robustness of uniform design}

\bigskip \noindent If the
experimenter is unable to make an informed choice of the assumed values for local optimality, the natural design choice is the uniform design ${\mathbf p}_{u}=(1/4,1/4,1/4,1/4).$ The relative loss of efficiency of ${\mathbf p}_{u}$ with respect to the true ${\mathbf w} = (w_1, w_2, w_3, w_4)$ is:
\begin{equation*}
R_{u}({\mathbf w})=\frac{\det \left( {\mathbf w}, {\mathbf p}_t\right) ^{1/3}-\det \left(
{\mathbf w},{\mathbf p}_{u}\right) ^{1/3}}{\det \left( {\mathbf w}, {\mathbf p}_{t}\right) ^{1/3}}
= 1 - \frac{1}{4}\left(\frac{ v_1+v_2+v_3+v_4}{L({\mathbf p}_t)}\right)^{1/3},
\end{equation*}%
where $v_i = 1/w_i$, ${\mathbf p}_t$ is the optimal design under ${\mathbf w}$,
and $L({\mathbf p}_t)$ is defined in (\ref{simplifiedprob}).
To study the maximum loss of efficiency, we consider $R_{\rm max}^{(u)}=\underset{{\mathbf w}}{\max }$ $R_{u}({\mathbf w})$. The following theorem finds the values of $R_{\rm max}^{(u)}$ for different values of ${\mathbf w}$'s.
\begin{theorem}\label{thm:42}
Without any loss of generality, assume $v_1\geq v_2\geq v_3\geq v_4$. Then \vspace{-.1in}
\[
R_{\rm max}^{(u)} = \left\{
\begin{array}{ll}
1 - \frac{3}{4}\left(1+\frac{3a}{b} \right)^{1/3} & \mbox{if\hspace{.1 in}} v_1 \geq v_2+v_3+v_4 \mbox{\hspace{.1 in}with\hspace{.1 in}} 0 < a\le v_i \le b\\
1 - \frac{3}{4}\times 2^{1/3} & \mbox{if\hspace{.1 in}} v_1 < v_2+v_3+v_4 \\
\end{array}
\right.
\]
\end{theorem}

\bigskip \noindent If $v_1 < v_2 + v_3 + v_4$, the maximum loss of the uniform design is $1 - \frac{3}{4}\times 2^{1/3} \approx 0.055 $ regardless of the choice of link functions. The uniform design clearly performs very well. Now, let us consider the saturation condition $v_1 \geq v_2 + v_3 + v_4$. For some reasonable choices of $\mathbf w$'s, Table~\ref{tab4.1} gives results for popular link functions. These $R_{\rm max}^{(u)}$'s are significantly larger than 0.055. Moreover, the efficiency of the uniform design gets worse as $a$ decreases to zero. This is intuitive because as $a$ decreases, $2v_{\rm max} - \sum v_i$ increases, worsening the performance of uniform design. Nevertheless, based on Theorem~\ref{thm:1}, if we \textit{know} that $v_{1}\geq v_{2}+v_{3}+v_{4}$ the design of choice clearly is the saturated design $(0,1/3,1/3,1/3).$ We will discuss the utility of saturated designs in practical applications in section~6. It can be seen from Theorem \ref{thm:42} that in general the maximum loss of efficiency is $1/4$.

\begin{table}[h]\caption{\textit{Maximum} loss of efficiency for $v_1 \geq v_2+v_3+v_4$}\label{tab4.1}
\begin{center}
\begin{tabular}{|c|c|c|}
\hline
    & \multicolumn{2}{c|}{Link functions}\\
    \cline{2-3}
    &  Logit                     & Probit or         \\%[-1pt]
    &                            & (complementary) log-log  \\
    &  $0.05 \le w \le 0.25$     & $0.05 \le w \le 0.65$ \\
      \hline
$R_{\rm max}^{(u)}$    & 0.123 & 0.196 \\      \hline
\end{tabular}
\end{center}
\end{table}

\vspace{-1in}

%\clearpage
\vspace{1in}
\setcounter{section}{1}
\setcounter{table}{0}
\setcounter{chapter}{5}
\setcounter{equation}{0} %-1
\bigskip \noindent {\bf 5. General case: $2^{k}$ experiment}

\bigskip \noindent The general case of $2^{k}$ factorial is technically complicated. We have some results for particular cases that will be reported in future publications. In this section we will show that the uniform design ($p_{1}=\cdots=p_{2^{k}}=\frac{1}{2^{k}}$) has a \textit{maximin} optimality property, i.e., it maximizes a lower bound of the {\it D-criterion}.

\bigskip \noindent Suppose there are $q$ parameters (main effects and interactions) in the chosen model for the $2^{k}$ experiment, i.e. $\beta $ is a $q\times 1$ vector and the design matrix $X$ is $2^{k}\times q.$ The matrix $X$ can be extended to the $2^{k}\times 2^{k}$ design matrix $F$ of the ``full model", i.e., one that includes all main effects and interactions. Note that $F$ is a Hadamard matrix, i.e., $F^{\prime }F=2^kI_{2^{k}}$, where $I_{2^k}$ is the identity matrix of order $2^{k}.$ Moreover, we can partition $F$ as
%\begin{equation*}
$F=\left( X|R\right)$
%\end{equation*}%
where $R$ contains all effects that are not in the chosen model. Clearly,%
\begin{equation*}
\left\vert F^{\prime }WF\right\vert =\left\vert X^{\prime }WX\right\vert
\left\vert R^{\prime }WR-R^{\prime }WX\left( X^{\prime }WX\right)
^{-1}X^{\prime }WR\right\vert \leqslant \left\vert X^{\prime }WX\right\vert
\left\vert R^{\prime }WR\right\vert .
\end{equation*}%
Hence \vspace{-.2in}
\begin{equation*}
\left\vert X^{\prime }WX\right\vert \geqslant \frac{\left\vert F^{\prime
}WF\right\vert }{\left\vert R^{\prime }WR\right\vert }.
\end{equation*}%
Note that $\left\vert F^{\prime }WF\right\vert =2^{kq}\Pi w_{i}\Pi p_{i},$ and
%\begin{equation*}
$
\left\vert R^{\prime }WR\right\vert \leqslant \left( w_{M}\right)
^{2^{k}-q}\left\vert R^{\prime }PR\right\vert
$
%\end{equation*}%
where $w_{M}=\max \{ w_{1},...,$ $w_{2^{k}}\}$, and $P=diag\left( p_{1},...,p_{2^{k}}\right)$. It follows from Proposition 1$'$ of Kiefer (1975) that the uniform design $p_{1}=\cdots=p_{2^{k}}=\frac{1}{2^{k}} $ maximizes $\left\vert R^{\prime }PR\right\vert $. Hence we obtain a \textit{lower bound} to the {\it D-criterion}%
\begin{equation*}
\left\vert X^{\prime }WX\right\vert \geqslant \frac{2^{k2^{k}}\Pi w_{i}\Pi
p_{i}}{\left( w_{M}\right) ^{2^{k}-q}}.
\end{equation*}%
This lower bound is a maximum when $p_{1}=\cdots=p_{2^{k}}=\frac{1}{2^{k}}$, which gives a maximin property of the uniform design.

\bigskip \noindent How good is the uniform design? The loss of efficiency corresponding to the uniform design is
\[
R_{u}(w)=1-\left( \frac{\left\vert X^{\prime }diag\left(
w_{1}/2^{k},...,w_{2^{k}}/2^{k}\right) X\right\vert }{\max \left\vert
X^{\prime }WX\right\vert }\right) ^{1/q}.
\]%
Then it can be shown that%
\begin{eqnarray*}
R_{u}(w) &\leqslant &1-\frac{\left\vert X^{\prime }W_{0}X\right\vert ^{1/q}}{%
2^{k}w_{M}} \leqslant 1-\frac{w_{m}}{w_{M}}
\end{eqnarray*}%
where $w_{m}=\min \left\{ w_{1},...,w_{2^{k}}\right\} $ and $%
W_{0}=diag\left( w_{1},...,w_{2^{k}}\right) .$

\bigskip \noindent For example, suppose it is known that $0.14\leqslant w_{i}\leqslant 0.20.$ Then the uniform design is not less than \thinspace $70\%$ efficient, irrespective of the value of $k$ or the $X$ matrix (the model). While this bound enables us to make general statements like this, the results in Theorem~\ref{thm:42} for the $2^{2}$ case shows that this bound can be quite conservative.

\setcounter{chapter}{6}
\setcounter{equation}{0} %-1
\bigskip\noindent {\bf 6. Discussion}

\bigskip \noindent Locally optimal designs require assumed values of the parameters $w$ which may not be readily available at the planning stage. The expression of $w_{i}$ given in section 2 depends on the assumed values of the parameter $\beta $ and the link function $\eta $. In situations where there is overdispersion, this expression (the nominal variance) may not be adequate to describe variation in the model, and a more realistic representation for $w$ may be \[ w_{i}=c_{i}\left( \frac{d\mu _{i}}{d\eta _{i}}\right) ^{2}/\left( \mu _{i}(1-\mu _{i})\right) \] where $c_{i}$ is a function of the factor levels $x_{1},x_{2},\ldots,x_{k}.$ This will make the specification of $w_{i}$ even more difficult. For the design problem, however, it can be seen that we need only the relative magnitudes $w_{i}^{\ast }=$ $w_{i}/w_{M},$ which may be easier to specify in some applications. Note that $w_{i}^{\ast }$ take values in the interval $(0,1].$

\bigskip \noindent The overall conclusion for the $2^{2}$ factorial experiment with main effects model is that, for the link functions we studied, the locally optimal designs are robust in the sense that the loss of efficiency due to misspecification of the assumed values is not large. If the (assumed) variance at one point is substantially larger than the others, then the D-optimal design is based on only $3$ of the $4$ points. In real world experiments, however, an experimenter would rarely feel confident to not allocate observations at a point based solely on assumed values, and this is not our recommendation for practice. However, the D-optimal design would still provide a useful benchmark for the efficiency of designs, and to the extent feasible it is wise to dedicate more resources to points that we believe have small variance and less resources to points with large variance. If there is no basis to make an informed choice of the assumed values, we \textit{can} recommend the use of the uniform design.

\bigskip \noindent We have extensive results on the robustness of the $2^2$ designs against the misspecification of $w$, which will be reported in another publication. This paper consists of initial results in this area of optimal designs for two level factorial experiments with binary response. Our research is ongoing on extending these results to more general factorial and fractional factorial experiments.

\bigskip \noindent {\bf Acknowledgment}

\noindent The authors thank Professor John P Morgan for his valuable suggestions. This research was supported by grants from the National Science Foundation.

\clearpage

\renewcommand{\theequation}{A.\arabic{equation}}
\appendix

\begin{center}\noindent {\large\bf Appendix}\end{center}

\noindent {\textbf{\emph{1. Proof of Theorem~\ref{thm:1}}}}

\smallskip\noindent
{\bf (1)} If $v_1\geq v_2+v_3+v_4$, then
\begin{eqnarray}
L &=& v_4(p_1p_2p_3+p_2p_3p_4)+v_3(p_1p_2p_4+p_2p_3p_4)+v_2(p_1p_3p_4+p_2p_3p_4) \nonumber\\
  & & +(v_1-v_2-v_3-v_4)p_2p_3p_4 \nonumber\\
  &=& v_4(p_1+p_4)p_2p_3+v_3(p_1+p_3)p_2p_4+v_2(p_1+p_2)p_3p_4\nonumber\\
  & & +(v_1-v_2-v_3-v_4)p_2p_3p_4\nonumber\\
  &\leq & v_4\left(\frac{(p_1+p_4)+p_2+p_3}{3}\right)^3 + v_3\left(\frac{(p_1+p_3)+p_2+p_4}{3}\right)^3\nonumber\\
  & & + v_2\left(\frac{(p_1+p_2)+p_3+p_4}{3}\right)^3 + (v_1-v_2-v_3-v_4)p_2p_3p_4\label{inequality1}\\
  &=& \frac{v_4}{27} + \frac{v_3}{27} + \frac{v_2}{27} + (v_1-v_2-v_3-v_4)p_2p_3p_4\nonumber\\
  &\leq & \frac{v_2+v_3+v_4}{27} + (v_1-v_2-v_3-v_4)\left(\frac{p_2+p_3+p_4}{3}\right)^3\label{inequality2}\\
  &\leq & \frac{v_2+v_3+v_4}{27} + (v_1-v_2-v_3-v_4)\left(\frac{p_1+p_2+p_3+p_4}{3}\right)^3\label{inequality3}\\
  &=& \frac{v_1}{27}\nonumber
\end{eqnarray}
By the inequality of arithmetic and geometric means,
the ``=" in (\ref{inequality1}) is true if and only if
$$p_1+p_4=p_2=p_3,\>\> p_1+p_3=p_2=p_4,\>\mbox{ and }\>
p_1+p_2=p_3=p_4$$
which implies $p_1=0$, $p_2=p_3=p_4=1/3$.
Note that the ``=" in (\ref{inequality2}) is true if
$p_2=p_3=p_4$, and the ``=" in (\ref{inequality3}) is true
if $p_1=0$.  After all, $L=v_1/27$ if and only
if $p_1=0$, $p_2=p_3=p_4=1/3$.

\smallskip\noindent
{\bf (2)} If ${\mathbf p} = (0, 1/3, 1/3, 1/3)$ maximizes (\ref{simplifiedprob}),
we claim that $v_1\geq v_2+v_3+v_4$. Otherwise, if $v_1 < v_2+v_3+v_4$, the solution ${\mathbf p}_\epsilon
=\left(\epsilon, (1-\epsilon)/3, (1-\epsilon)/3, (1-\epsilon)/3\right)$
will be better than ${\mathbf p}$ for small enough $\epsilon > 0$.
It can be shown that
$$L\left({\mathbf p}_\epsilon\right) > L({\mathbf p})
\mbox{ \hspace{.2in} if and only if  \hspace{.2in}  } 3 d_1 > \epsilon (3 + 6 d_1 - 2 \epsilon - 3 d_1 \epsilon)
$$
where $d_1 = (v_2 + v_3 + v_4 - v_1)/v_1 > 0$.
\hfill{$\Box$}

\bigskip \noindent {\textbf{\emph{2. Proof of Theorem~\ref{thm:42}}}}

\bigskip \noindent
To maximize $R_u({\mathbf w})$ is equivalent to minimize
$$Q(v_1, v_2, v_3, v_4) = \frac{v_1+v_2+v_3+v_4}{v_1 p_2 p_3 p_4
 + v_2 p_1 p_3 p_4 + v_3 p_1 p_2 p_4 + v_4 p_1 p_2 p_3}~.
$$
where $(p_1, p_2, p_3, p_4)$ is the optimal allocation for the given  $(v_1, v_2, v_3, v_4)$.   Note that $v_1\geq v_2\geq v_3\geq v_4$ implies that $p_1 \leq p_2 \leq p_3 \leq p_4$.

\smallskip\noindent
{\bf (1)}
If $v_1 \geq v_2 + v_3 + v_4$, then $p_1 = 0$ and $p_2=p_3=p_4=1/3$ which gives
$$R_u({\mathbf w}) = 1 - \frac{3}{4}\left(1 + \frac{v_2 + v_3 + v_4}{v_1}\right)^{1/3}.$$
Suppose $0< a \leq v_i \leq b$ for $i=1,\ldots,4$, then $v_1 \geq v_2 + v_3 + v_4$ implies $b\geq 3a$ and \vspace{-.1in}
$$Q(v_1,v_2,v_3,v_4) = 27\left[1+\frac{v_2+v_3+v_4}{v_1}\right]
\geq 27\left[1+\frac{3a}{b}\right].$$
The minimum of $Q$ is attained at $v_1 =b$ and $v_2=v_3=v_4=a$.
So $R_{\max}^{(u)} =1 - \frac{3}{4}(1 + 3a/b)^{1/3}$.

\vspace{-.2in}

\begin{equation}\label{thm42eq}
\mbox{Specifically, if $v_1 = v_2 + v_3 + v_4$, $Q(v_1, v_2, v_3, v_4) = 54$ and $R_u({\mathbf w}) = 1 - \frac{3}{4}\times 2^{1/3}$}.
\end{equation}

\bigskip \noindent
{\bf (2)} If $v_1 < v_2 + v_3 + v_4$, then $p_1 > 0$.
Let $\delta > 0$ be small enough so that $p_1'=p_1 - \delta (1-p_1) > 0$.
Let $p_i' = (1 + \delta)p_i$, $i = 2, 3, 4$. Then $p_1' + p_2' + p_3' + p_4'=1$.
It can be verified that for any $v_1' > v_1$,
$$
Q(v_1', v_2, v_3, v_4) \leq \frac{v_1'+v_2+v_3+v_4}{v_1' p_2' p_3' p_4'
 + v_2 p_1' p_3' p_4' + v_3 p_1' p_2' p_4' + v_4 p_1' p_2' p_3'}
 < Q(v_1, v_2, v_3, v_4)
$$
for small enough $\delta>0$. On the other hand, it can be verified that \vspace{-.1in}
$$\lim_{v_1 \uparrow (v_2 + v_3 + v_4)} Q(v_1, v_2, v_3, v_4)
= Q(v_2 + v_3 + v_4, v_2, v_3, v_4) = 54$$
regardless of the values of  $v_2, v_3, v_4$.  So by (\ref{thm42eq})
the maximum of $R_u({\mathbf w})$ is $1 - \frac{3}{4}\times 2^{1/3}$ under the restriction
$v_1 < v_2 + v_3 + v_4$. \hfill{$\Box$}

\clearpage
\singlespacing

\noindent{\large\bf References}
\begin{description}

\item Chernoff, H. (1953). Locally optimal designs for estimating parameters. {\em Annals of Mathematical Statistics}, {\bf 24}, 586$-$602.

%\item Collett, D. (1991). {\em Modelling Binary Data}, Chapman \& Hall/CRC, New York.

\item Crowder, M. J. (1978). Beta-binomial anova for proportions, \emph{Applied Statistics}, {\bf 27}, 34$-$37.

\item Fotiou, I. A., Parrilo, P. A., and Morari, M. (2005). Nonlinear parametric optimization using cylindrical algebraic decomposition. In \emph{Proceedings of the 44th IEEE Conference on Decision and Control, and the European Control Conference 2005}, 3735$-$3740.

\item Hamada, M. and Nelder, J. A. (1997). Generalized linear models for quality-improvement experiments, \emph{Journal of Quality Technology}, \textbf{29}, 292$-$304.

\item Hoblyn, T. N. and Palmer, R. C. (1934). A complex experiment in the propagation of plum rootstocks from root cuttings: season, 1931$-$1932, {\em Journal of Pomology and Horticultural Sciences}, {\bf 12}, 36$-$56.

\item Karush, W. (1939). Minima of functions of several variables with inequalities as side constraints. M.Sc. Dissertation. Dept. of Mathematics, Univ. of Chicago, Chicago, Illinois.

\item Kiefer, J. (1975). Construction and optimality of generalized Youden designs. In: J.N. Srivastava, Ed., {\em A Survey of Statistical Design and Linear Models}, North-Holland, Amsterdam, 333$-$353.

\item Khuri, A. I., Mukherjee, B., Sinha, B. K. and Ghosh, M. (2006). Design Issues for Generalized Linear Models: A Review. {\em Statistical Science}, {\bf 21}, 376$-$399.

\item Kuhn, H. W. and Tucker, A. W. (1951). Nonlinear programming. {\em Proceedings of 2nd Berkeley Symposium}, Berkeley: University of California Press, 481$-$492.

\item Martin, B., Parker, D. and Zenick, L. (1987). Minimize slugging by optimizing controllable factors on topaz windshield molding, In: \emph{Fifth Symposium on Taguchi Methods, American Supplier Institute, Inc., Dearborn, MI}, 519$-$526.

\item McCullagh, P. and Nelder, J. (1989). {\rm Generalized Linear Models}, Second Edition, Chapman and Hall/CRC, Boca Raton.

\item Myers, R. M., Montgomery, D. C. and Vining, G. G. (2002). {\em Generalized Linear Models with Applications in Engineering and Statistics}, John Wiley, New York.

\item Rao, C.R. (1947). Factorial experiments derivable from combinatorial arrangements of arrays. \emph{Journal of the Royal Statistical Society}, {\bf 9},  128$-$139.

\item Rao, C. R. (1971). Unified theory of linear estimation (Corr: 72V34 p194; 72V34 p477). {\rm Sankhy\={a}, Series A, Indian Journal of Statistics}, {\bf 33}, 371$-$394.

\item Smith, W. (1932). The titration of antipneumococcus serum, {\em Journal of Pathology}, {\bf 35}, 509$-$526.

\end{description}

\clearpage

\end{document}